\documentclass[a4paper,12pt]{article}
\pdfoutput=1
\usepackage{hyperref}
\usepackage{epsfig}
\usepackage{amssymb}
\usepackage{cancel}
\usepackage{setspace}
\usepackage[usenames,dvipsnames,svgnames,table]{xcolor}
\usepackage{amsmath}
\usepackage{graphicx}
\usepackage{subfigure}
\usepackage{url}
\usepackage{slashed}
\usepackage{booktabs}
\usepackage{cite}
\def\eq#1{{Eq.~(\ref{#1})}}
\def\eqs#1#2{{Eqs.~(\ref{#1})--(\ref{#2})}}

\newcommand{\gsim}{\lower1.0ex\hbox{$\;\stackrel{\textstyle>}{\sim}\;$}}
\newcommand{\lsim}{\lower1.0ex\hbox{$\;\stackrel{\textstyle<}{\sim}\;$}}

\definecolor{oucrimsonred}{rgb}{0.6, 0.0, 0.0}
\definecolor{persianblue}{rgb}{0.11, 0.22, 0.73}
\definecolor{forestgreen}{rgb}{0.13,0.35,0.13}
\def\hhref#1{\href{http://arxiv.org/abs/#1}{#1}} 
\newcommand{\be}{\begin{equation}}
\newcommand{\ee}{\end{equation}}
\newcommand{\bea}{\begin{eqnarray}}
\newcommand{\eea}{\end{eqnarray}}
\newcommand{\nn}{\nonumber}
\newcommand{\dph}{\ensuremath{\bar{\gamma}}}

\def\G{\scriptscriptstyle{G}}

\def\eq#1{{Eq.~(\ref{#1})}}
\def\eqs#1#2{{Eqs.~(\ref{#1})--(\ref{#2})}}

\begin{document}

\begin{center}
  {\Large  {\bf Untangling the spin of a dark boson  in $Z$ decays}}

\vspace{1.5cm}
{\bf Andrea Comelato$^{a}$}
and
{\bf Emidio Gabrielli$^{a,b,c}$}

\vspace{0.5cm}
{\small       {\it
          $^{a}$Department of Physics - Theoretical Section, University of Trieste\\ Strada Costiera 11 - 34151, Trieste, Italy\\
          $^{b}$INFN, Sezione di Trieste, Via  Valerio 2, 34127 Trieste, Italy\\
         $^{c}$NICPB, R\"avala 10, Tallinn 10143, Estonia}
\\[1mm]}
       
\vspace*{2cm}{\bf ABSTRACT}
\end{center} 
\vspace{0.3cm}

We analyze the $Z$-boson decay $Z\to \gamma\, X$  into a photon ($\gamma$) plus a hypothetical light boson ($X$) belonging to a dark or secluded sector. 
Due to its feeble interactions with Standard Model fields, this dark boson is behaving as missing energy in the detector. We consider for $X$ the cases of  spin-1 (massless dark-photon), spin-0 (axion-like), and spin-2 (graviton-like) particles and explore the way to untangle its spin origin. All these scenarios predict a universal signature for this decay, characterized by a single mono-chromatic photon in the $Z$ center of mass, with energy about half of the $Z$ mass, plus a neutrino-like missing energy associated to the $X$ boson.
We show that if the $Z\to \gamma\, X$  signal should be discovered  at $e^+e^-$ colliders, the angular distribution of the mono-chromatic photon in $e^+e^-\to Z\to \gamma\, X$ can provide a clean probe to discriminate between the $J=1$ and alternative $J=0/2$ spin nature of the dark boson.

\newpage
\section{Introduction}
\label{sec:intro}
The lack of any experimental evidence at the LHC for a heavy New Physics (NP) above the TeV scale \cite{Atlas:2019qfx}, as expected by the many NP scenarios beyond the Standard Model (SM) theory, is changing our perspective about the search for a NP. The accessible sector of NP could be instead made up of light new particles, feebly coupled to SM fields, as predicted by scenarios with dark or secluded sectors beyond the SM, where for instance the candidate(s) for dark matter might reside. The dark sector, consisting of new particles which are singlet under the SM gauge interactions, can indeed have its own long range interactions, characterized by massless or very light mediators, like the dark-photon, the quantum field associated to a $U(1)_D$ gauge invariance in the dark sector.
These scenarios have motivated the search for weakly coupled light particles,    as it can be seen by the many theoretical and experimental works on this subject \cite{Raggi:2015yfk}.

In this framework, we focus on the effective couplings of a light and long-lived neutral $X$ boson with the neutral sector of electroweak gauge bosons of the SM. In particular, we explore, in a model independent way, the production of $X$ by means of the $Z$ boson decay into
\bea
Z\to \gamma\, X\, ,
\label{Zdecay} 
\eea
where $X$ it is assumed to behave as missing energy in the detector.

The striking experimental signature of this decay, in the $Z$ rest frame, is then characterized by an isolated mono-chromatic photon, with energy (almost) half of the $Z$ mass , and missing energy with (almost) vanishing invariant mass for a massless (massive) $X$. 

The best place to look for the process in \eq{Zdecay} is at $e^+e^-$ colliders, where the main characteristic of the signature is maintained, although the mono-chromaticity of the photon is slightly spread by the initial \textit{Bremsstrahlung} radiation. Moreover, rare $Z$ decays are expected to be investigated at the  Future Circular Collider (FCC-ee), with its projected production of $10^{13}~Z$ bosons\cite{Abada:2019zxq}.
This process was already explored at the experimental level at the Large Electron-Positron Collider (LEP) via
\be e^{+} e^{-} \to Z \to \gamma + X\, ,
\label{eeZdecay}
\ee
where $X$ stands for no other detected neutral particles. Negative evidence for this signal, set a limit of $10^{-6}$ at the 95\% CL on the corresponding branching ratio (BR), in the case of a massless final state $X$ \cite{Acciarri:1997im}. On the other hand, at hadron colliders this signal would be rather difficult to detect, due to the challenging reconstruction of the $Z$ invariant mass and the large background of soft jets faking the missing energy.

This process has been recently analyzed in the case of $X$ as a massless dark-photon\cite{Fabbrichesi:2017zsc}. 
The dark photon scenario has been extensively analyzed in the literature, mainly for the massive case, and it is also the subject of many current experimental searches, see \cite{Alexander:2016aln} and \cite{Fabbrichesi:2020wbt} for a more recent review. Most of the experimental searches focus on massive dark photons, where the $U(1)_D$ gauge field generates, through a potential kinetic mixing with the photon, a tree-level (milli-charged) interactions with ordinary charged SM particles. On the other hand, for a massless dark-photon the kinetic mixing can be fully rotated away leading to  dark-photon interactions with ordinary matter mediated by effective higher-dimensional operators\cite{Holdom:1985ag}. The leading coupling of a massless dark photon to SM charged particles is provided by the magnetic- and electric-dipole interactions\cite{Holdom:1985ag,Dobrescu:2004wz}, including the flavor-changing ones \cite{Gabrielli:2016cut}. 
Phenomenological implications of massless dark-photon scenarios have been recently explored in the framework of Higgs boson \cite{Biswas:2016jsh} and rare Kaon decays \cite{Fabbrichesi:2017vma}.

Recently, in \cite{Fabbrichesi:2017zsc}  it has been shown that the $Z$ can decay at 1-loop into a photon and massless dark photon without violating the Landau-Yang theorem~\cite{Landau:1948kw}, due to the fact that the dark and the ordinary photon are distinguishable particles. An upper limit on the viable BR for the decay $Z\to \gamma \dph$ has been estimated to be of the order of $O(10^{-9})$ \cite{Fabbrichesi:2017zsc}, in the framework of a simplified model of the dark sector. These results hold also for a massive dark-photon, due to its own magnetic-dipole interactions with SM fields.\footnote{The $Z\to \gamma V$ decay has been also explored in \cite{Dror:2017ehi} for a massive vectorial field $V$ (not exactly a dark-photon) coupled to anomalous non-conserved $U(1)$ currents, via gauging anomalous $U(1)$ symmetries of the SM \cite{Dror:2017ehi}.}

We will explore here the possibility that other $X$ spin configurations can mimic the same signature of a massless dark photon in \eq{Zdecay}, and  show how to disentangle a genuine spin-1 dark-photon signal against possible $X$ candidates of different integer spin. We will assume an uncertainty of the order of a $1{\rm GeV}$ in the invariant mass of the missing energy, mainly due to the detector performance in the reconstruction of the missing mass. Therefore, in alternative to the massless dark-photon, we consider at phenomenological level the hypothetical scenarios of spin-0 and spin-2 particles  with masses $m_X$ below the $1{\rm GeV}$ scale, which are inspired by known theoretical frameworks.

In this respect, we consider first, as an alternative to the dark photon, $X$ to be a light axion-like particle (ALP), in both scalar and pseudoscalar scenarios. 
The ALPs have been predicted in several SM extensions, mainly motivated by the solution to the strong-CP problem, where the ALP is a QCD axion \cite{Peccei:1977hh}), or being associated to pseudo-Nambu-Goldstones bosons corresponding to spontaneously broken continuous symmetries (either in the visible or dark sector), as well as to a moduli field in string models \cite{Witten:1984dg,Svrcek:2006yi,Arvanitaki:2009fg,Acharya:2010zx}. The phenomenological aspects of the ALPs have been extensively investigated in recent years, especially collider search of ALP's ~\cite{Bauer:2018uxu,Bauer:2017ris,Dolan:2017osp}. The most sever constraints on the ALP couplings are in the range of masses below the MeV scale, mainly due to low energy observables and constraints from astrophysics and cosmology ~\cite{Bauer:2017ris}.

The $Z$ decay process in \eq{Zdecay}, with $X=a$ as ALP, has been considered in the literature \cite{Kim:1989xj}, More recently, a dedicated study on the sensitivity of the $Z\to a \gamma$ decay at the LHC and future colliders, via visible ALP decays into two photons and/or lepton pairs~\cite{Bauer:2018uxu,Bauer:2017ris} has been explored for various range of masses.
Present constraints on the effective scale, mainly based on previous LEP analysis on ALP visible decays, allows for BR($Z\to a \gamma)$ as large as $O(10^{-4})$ for the range of masses $100 {\rm MeV} \lsim m_a \lsim 1{\rm GeV}$.
We will show that, under the requirement for the ALP to behave as missing energy in the detector, stronger constraints on the BR for this decay apply, that could reach $O(10^{-6})$ for masses of the order of $1{\rm GeV}$. This is also consistent with the LEP bound \cite{Acciarri:1997im} that applies to the corresponding signature. These results can be easily generalized to ALP in both scalar and pseudoscalar cases. Therefore, a large number of viable events for $Z\to a \gamma$ at future $e^+e^-$ colliders with high luminosity are expected, that could be competitive with the corresponding ones from the $Z\to \gamma \bar{\gamma}$ signature.

Next, we consider a more exotic scenario for $X$ as a ultralight massive spin-2 particle $G$. Fundamental massive spin-2 fields have been predicted by several extensions of gravity theories, like the massive Kaluza-Klein (KK) excitations of the standard massless graviton in quantum gravity (QG) theories in large extra-dimensions (ADD \cite{ArkaniHamed:1998rs} and RS \cite{Randall:1999ee} scenarios), as well as the massive graviton in the bi-metric theories \cite{Schmidt-May:2015vnx,Hinterbichler:2011tt,deRham:2014zqa}.
For the purposes of the present analysis, we do not make any assumption about the origin of this field. Since we are only interested in the phenomenological implications of $Z\to \gamma G$ decay, we restrict the analysis to the effects of the linear theory (with an on-shell $G$ field as external source) in flat space-time, common characteristic to many extended gravity scenarios. By consistency we assume the spin-2 field to be universally coupled to the energy-momentum tensor of SM fields, as for the linear  graviton-like coupling to SM fields, with an effective scale $\Lambda_G$. Then, the effective $Z \gamma G$ vertex is predicted as a function of $\Lambda_G$ to be finite, induced at 1-loop by SM fields running as virtual particles.

In order to avoid constraints from short-range gravity experiments (see \cite{Murata:2014nra} for a recent review) and mimic a neutrino-like signature, we restrict its mass to lie the range ${\rm eV} \lsim m_G \lsim  1 {\rm GeV}$, with an effective scale $\Lambda_G \ge {\rm TeV}$, and require that it does not decay inside the detector. We will show, that for a spin-2 particle subject to these specific constraints, predictions for ${\rm BR}(Z\to \gamma G)$ as large as $O(10^{-8})$ are possible, thus in the sensitivity range considered here for the $Z\to \gamma X$.
\footnote{The decay in \eq{Zdecay} with a massive spin-2 $X$ has been analyzed in~\cite{Allanach:2007ea} in the framework of the ADD scenario \cite{ArkaniHamed:1998rs}, predicting a viable BR of the order of $O(10^{-11})$ for D=2. However, there the signature is different from the one analyzed here, due to the absence of events with a mono-chromatic photon characteristic. Indeed, in ADD this decay can only be observed as inclusive production of a (almost) continuum spectrum of KK graviton excitations, behaving as missing energy, thus reflected in a (almost) continuum photon spectrum.}

Now, assuming the process in \eq{Zdecay} will be observed with a BR in the sensitivity range of ${\rm BR}(Z\to \gamma X)\sim 10^{-12}-10^{-6}$, given the possibility that $X$ might belong to one of these scenarios, one may wonder if its spin nature could be disentangled by analyzing the angular distributions of the outgoing photon. Clearly, the answer is not, if the $Z$ boson is unpolarized.  Indeed, in the unpolarized $Z\to \gamma X$ decay the photon will be isotropically distributed, independently on the spin nature of the $X$ particle. However, a non-trivial angular distribution of the photon, that depends on the $X$ spin, can appear in the case of polarized $Z$ decays. Remarkably, one of the main features of the $e^+e^-$ colliders at the resonant $Z$ peak, is that the on-shell $Z$ boson is always produced polarized, thus transmitting the $Z$-spin correlations to the final state. In this regard, we will show that the angular distribution of the mono-chromatic photon in the $e^+e^-\to Z\to \gamma\, X$ process at the $Z$ peak can offer a clean probe to untangle the spin-1 nature of the $X$ boson against other possible spin-0/2 interpretations.

The paper is organized as follows. In section 2 we will give the expressions for the effective Lagrangians relevant to the decay $Z\to \gamma X$  for the three spin scenarios mentioned above, providing the corresponding amplitudes and total rates, as well as a discussion on the corresponding allowed range of branching ratios. In section 3 we analyze the angular distributions of polarized $Z$ decays in each spin $X$ scenario, while the corresponding results for a $Z$ produced in resonant $s$-channel at $e^+e^-$ colliders will be presented in section 4. Finally, our conclusions are reported in section 5.

\section{Effective Lagrangians and Amplitudes}
\subsection{Spin-1: massless dark photon}
\label{sec:DP}
We consider here the case of $X$ as a massless dark-photon $\bar\gamma$, which is effectively coupled to the photon $\gamma$ and $Z$ gauge boson. Generalization to the massive dark-photon in the limit of small mass are straightforward. We recall first the main results obtained in \cite{Fabbrichesi:2017zsc}.

The lowest dimensional gauge-invariant Lagrangian (CP even) for the leading contribution to the effective $Z\gamma\bar{\gamma}$ vertex, has been derived in \cite{Fabbrichesi:2017zsc}. We parametrize this Lagrangian as
\bea
{\cal L}_{eff}= \frac{e}{\Lambda M_Z}\sum_{i=1}^3 C_i {\cal O}_i(x) \, ,
\label{LeffMD}
\eea
where $e$ is the unit of electric charge, $\Lambda$ is the scale of the new physics, the dimension-six operators ${\cal O}_i$ are given by
\bea
{\cal O}_1 (x )&=&Z_{\mu\nu}\tilde{B}^{\mu\alpha} A^{\nu}_{~\alpha} \, ,\\
{\cal O}_2 (x) &=&Z_{\mu\nu}B^{\mu\alpha} \tilde{A}^{\nu}_{~\alpha} \, ,\\  
{\cal O}_3  (x) &=& \tilde{Z}_{\mu\nu}B^{\mu\alpha} A^{\nu}_{~\alpha}  \, ,
\eea
the field strengths $F_{\mu\nu}\equiv\partial_{\mu}F_{\nu} -\partial_{\nu}F_{\mu}$, for $F_{\mu\nu}=(Z,B,A)_{\mu\nu}$, correspond to the $Z$-boson ($Z_{\mu}$), dark-photon ($B_{\mu}$) and photon ($A_{\mu}$)  fields, respectively, and 
$\tilde{F}^{\mu\nu}\equiv \varepsilon^{\mu\nu\alpha\beta}F_{\alpha\beta}$ is the dual field strength. The expression for the coefficients  $C_M$ in \eq{dipole}, derived in \cite{Fabbrichesi:2017zsc}, can be found  in Appendix.

As mentioned in the introduction, the Landau-Yang theorem ~\cite{Landau:1948kw} can be avoided in the $Z\to \gamma \bar{\gamma}$ due the fact that the photon and the massless dark-photon are distinguishable particles. Less obvious is how this effective vertex can be generated from a UV theory. In \cite{Fabbrichesi:2017zsc} it has been demonstrated that the above Lagrangian in \eq{LeffMD} arises at low energy as an effective 1-loop contribution, with SM fermions running in the loop, because the dark-photon does not have tree-level couplings with SM fields. Indeed, the leading coupling of a massless dark-photon to charged SM fermions is via magnetic- or electric-dipole operators, namely
\be
{\cal L}_{\rm dipole}=\sum_f \frac{e_D}{2\Lambda} \bar{\psi}_f \sigma_{\mu\nu} \left( d^f_M + i \gamma_5 d^f_E \right)\psi_f 
B^{\mu\nu}\, ,
\label{dipole}
\ee
where $B^{\mu\nu}$ is the corresponding $U(1)_D$ field strength of dark photon field, the sum runs over all the SM fields, $e_D$ is the $U_D(1)$ dark elementary charge (we assume universal couplings), $\Lambda$ the effective scale of the dark sector, and $\psi_f$ a generic SM fermion field. The scale $\Lambda$ appearing in \eq{LeffMD} is the same of \eq{dipole}. The magnetic- and electric-dipole coefficients $d^f_M$  and $d^f_E$ respectively, can be computed from a renormalizable UV completion theory for the dark sector \cite{Fabbrichesi:2017zsc}.

If the dark-photon would have been coupled at tree-level with SM charged fermions (as for the ordinary photon or for the milli-charge couplings of massive dark-photon), the loop contribution would have been zero for each fermion running in the loop, in agreement with what is expected by the Landau-Yang theorem. Therefore, from the point of view of a renormalizable UV completion of the theory, the effective Lagrangian in \eq{LeffMD} is the result of a 2-loop effect, including the effective dipole interactions that originate from 1-loop \cite{Fabbrichesi:2017zsc}. The same conclusions hold for the massive dark-photon, since the effective $Z\gamma\bar{\gamma}$ can be induced by its own dipole-type of interactions as in \eq{dipole}.

Analogously, the CP-odd Lagrangian induced by the electric-dipole moment is instead 
\bea
{\cal L}^{(E)}_{eff}= \frac{e}{\Lambda M_Z}C_E {\cal O}(x)  \, ,
\label{LeffED}
\eea
where the dimension-six operator is 
\bea
{\cal O} (x)=Z_{\mu\nu} A^{\mu\alpha} B^{\nu}_{~\alpha} \, .
\eea
The expression for the coefficients  $C_E$ in \eq{dipole} is reported in Appendix and in \cite{Fabbrichesi:2017zsc}. The operators in \eq{LeffMD} and \eq{LeffED} are  $CP$ even and odd respectively.

Concerning the decay $Z\to \gamma \bar{\gamma}$, the corresponding amplitudes in momentum space can be found in \cite{Fabbrichesi:2017zsc}. Finally, by taking into accounts the effective Lagrangians in \eq{LeffMD} and \eq{LeffED} the
total width for the unpolarized $Z$ decay is given by
\bea
\Gamma (Z\rightarrow \gamma \bar \gamma) &=&  \frac{\alpha M^3_Z }{6 \Lambda^2}
\left(|C_{M}|^2 +  |C_{E}|^2\right) ,
\label{GammaZDP}
\eea
where  $C_{M}=\sum_i C_i$. Same results hold for the massive dark photon in the massless limit, with the scale $\Lambda$ corresponding to its dipole-interactions in \eq{dipole}.

As discussed in \cite{Fabbrichesi:2017zsc}, in the framework of a UV complete model for the dark sector, responsible to generate at 1-loop the dipole interactions in \eq{dipole}, it has been estimated that the largest allowed values for the BR could lie between ${\rm BR}(Z\to \gamma \bar{\gamma})\sim 10^{-11}$ and
${\rm BR}(Z\to \gamma \bar{\gamma})\sim 10^{-9}$, depending on the values of $\alpha_D$, the $U(1)_D$ coupling in the dark sector, and the $d_{M,E}^f$ couplings in the dipole-type of interactions in \eq{dipole}. However, these upper limits could be relaxed if a non-perturbative dynamics is responsible for these couplings potentially pushing up the BR close to the LEP upper bound of  ${\rm BR}(Z\to \gamma \bar{\gamma})\simeq 10^{-6}$.

As mentioned in the introduction, the best place to study this kind of signature is at the  $e^+e^-$ colliders. In particular, these BRs are in the ballpark of sensitivity of future $Z$ factories at $e^+e^-$ colliders, like for example the FCC-ee colliders\cite{Abada:2019zxq}. Assuming a collected number $N_Z=10^{13}$ of $Z$ boson events at the FCC-ee,  an expected $10^2-10^4$ number of  $Z\to \gamma \bar{\gamma}$ events would be possible, depending on the dark sector couplings.

\subsection{ Spin-0: ALP scalar and pseudoscalar}
\label{sec:ALP}
Here we consider a scenario for $X$ as an axion like particle (ALP), that can mimic the $Z\to \gamma X$ signature of a massless or ultralight dark photon.
We consider both the scenarios for $X$ as massive scalar $\varphi_S$ and pseudoscalar $\varphi_P$ particles and require them to behave as missing energy in the detector.

Let assume that this process is induced by a UV physics well above the EW scale. In this case an effective low energy Lagrangian approach can be used.
Then, we can parametrize the gauge-invariant contribution of the lowest dimensional operators of (dimension 5) to the corresponding  effective Lagrangians as
\bea
{\cal L}_{eff}^S&=& \frac{1}{\Lambda_S} \varphi_S Z_{\mu\nu} F^{\mu\nu}\\
{\cal L}_{eff}^P&=& \frac{1}{\Lambda_P} \varphi_P Z_{\mu\nu} \tilde{F}^{\mu\nu}\, ,
\label{Lscalar}
\eea
where $\Lambda_{S,P}$ are the corresponding effective scales.

Using the Lagrangians in \eq{Lscalar} the corresponding amplitudes $M_S$ ($M_P$) for the $Z$ decay into scalar (pseudoscalar) plus photon channel are
\bea
Z(p)&\to& \gamma(k)~ \varphi_A(q)
\eea
with $A=S,P$ are given by 
\bea
M_S&=& \frac{i}{\Lambda_S}\varepsilon_Z^{\mu}(p)\varepsilon^{\nu\star}(k)
\hat{T}^S_{\mu\nu}(p,k)\, ,
\nonumber\\
M_P&=& \frac{i}{\Lambda_P}\varepsilon_Z^{\mu}(p)\varepsilon^{\nu\star}(k)
\hat{T}^P_{\mu\nu}(p,k)\, ,
\eea
where $\hat{T}^S_{\mu\nu}(p,k)=2\left(\eta_{\mu\nu}(p\cdot k) -k_{\mu}p_{\nu}\right)$
and $\hat{T}^P_{\mu\nu}(p,k)=4\epsilon_{\mu\nu\alpha\beta}p^{\alpha}k^{\beta}$,
with $\eta_{\mu\nu}$ the Minkowski metric and $\epsilon_{\mu\nu\alpha\beta}$ the complete antisymmetric tensor.
Then, the corresponding total decay widths in the $Z$ rest frame is
\bea
\hat{\Gamma}_A&\equiv& \hat{\Gamma}(Z\to \gamma \varphi_A)\,=\,\frac{C_A M_Z^3}{24 \pi \Lambda_{S}^2}\left(1-r_A \right)^3\, ,
\label{GammaZS}
\eea
with $A=S,P$, where $C_S=1$ and $C_P=4$, and $r_A=m_A^2/M_Z^2$, with $m_A$ the mass of the scalar or pseudoscalar particle.

Now we consider some phenomenological implications of these results, in order to get a feeling with the expected BRs for the $Z\to \gamma \varphi_{A}$ decays.
If we assume the interactions in \eq{Lscalar}, then the ALP is a stable particle and automatically satisfies the missing-energy signature.
However, we conservatively consider a more realistic scenario, which is more theoretically justified. In particular, we assume the ALP to be effectively coupled, in addition to \eq{Lscalar}, two photons with the same strength as in \eq{Lscalar}, and require that it decays (in two photons) outside the detector. 

Let us focus only on the scalar case, since the pseudoscalar scenario should give comparable bounds.
At this aim, we consider in addition to \eq{Lscalar}, the existence of a new effective coupling to two photons in the Langrangian as 
\bea
    {\cal L}_{eff}^S&\supset & \frac{1}{\Lambda^{\gamma\gamma}_S} \varphi F_{\mu\nu} F^{\mu\nu}\, .
    \label{LSgg}
\eea
The reason to consider also the two photon interaction is that, from the point of view of a UV completion of the theory, one cannot avoid the presence of this interaction, if the $Z\gamma\varphi_S$ coupling in \eq{Lscalar} is present. Indeed, after the rotation into EW mass eigenstates, the two scales $\Lambda^{\gamma\gamma}_S$ and $\Lambda_S$ can be linearly related by coefficients proportional to the cosine and sine of the Weinberg angle ${\theta}_W$~\cite{Bauer:2018uxu,Bauer:2017ris}.
Then, a part from special UV models where one of the two couplings is tuned to cancel or be suppressed, these two scales are expected to be of the same order. The same conclusion does not hold for the Yukawa-like coupling of the ALP to fermions, with respect to the effective interactions in Eqs.(\ref{Lscalar}),(\ref{LSgg}), where these two different kind of interactions could be really independent from each other.\footnote{As an example, notice that the effective scales in Eqs.({\ref{Lscalar},\ref{LSgg}}) could be generated also in the absence of Yukawa couplings of the ALP to SM fermions, induced for instance by new heavy messenger scalar fields (EW charged and with trilinear couplings to ALP) running in the loop.} In order to stick on the most simple but non-trivial scenario, we assume the ALP couplings to fermions vanishing or being strongly suppressed, thus not contributing to the total width.
Then, since we are interested in the order of magnitude constraints on the effective scale $\Lambda_S$, we assume for simplicity $\Lambda^{\gamma\gamma}_S\sim \Lambda_S$, and set to zero all other ALP couplings to SM fields.

Under this setup, we can now analyze the constraints on the scalar or pseudoscalar mass against the corresponding effective scale $\Lambda$, that come from the requirement that the ALP does not decay inside the detector. Following the above considerations, total width of a scalar $X$ as ALP is given by
\bea
\hat{\Gamma}(S\to \gamma \gamma)&=& \frac{m_S^3}{16\pi\Lambda_S^2}\, ,
\eea
where $m_S$ is the mass of the scalar ALP. 

By requiring that the ALP does not decay inside the detector, that we conservatively take of length $L=10$m for $e^+e^-$ colliders, and assuming $\hat{\Gamma}(S\to \gamma \gamma)$ as the total width of ALP, we get
\bea
\Lambda_S \gsim  47 \left(\frac{m_S}{\rm 100\,MeV}\right)^2\, {\rm TeV}\, .
\label{limitLambdaS}
\eea

However, for masses below  $m_S < 100\, {\rm MeV}$, stronger limits on the effective scale  $\Lambda_S$ from astrophysics and low energy experiments apply, that are of the order of $\Lambda_S > 10^5 -10^6 {\rm TeV}$~\cite{Bauer:2018uxu,Bauer:2017ris}. These can largely overseed the bounds in \eq{limitLambdaS} with stronger constraints on $\Lambda_S$. These lower bounds imply ${\rm BR}(Z\to \varphi \gamma) < 10^{-13} (10^{-16}) $, corresponding to $\Lambda_S > 10^5 (10^6) {\rm TeV}$ respectively. As we can see, these BRs are too small to be detected, even for the high statistics of $Z$ that could be produced at the future FCC-ee collider.

Finally, we consider the next range of $m_S$ masses, namely from 100 MeV up to the ${\cal O}({\rm 1 GeV})$, where the kinematic properties of a neutrino-like $X$ signature might still hold, assuming the detector uncertainties does not allow to resolve $X$ masses below $1{\rm GeV}$. In this range of masses, there is still an unconstrained region of the effective scale $\Lambda_{S,P}$ from the searches at the Large Electron-Positron (LEP) collider, leaving to allowed values of the order of $\Lambda_{S,P} > {\cal O}(1{\rm TeV})$ scale~\cite{Bauer:2018uxu,Bauer:2017ris}, that would imply a viable BR of the order ${\rm BR}(Z\to \varphi \gamma) \sim 10^{-4}$. However, these constraints hold under the assumption that the scale $\Lambda_{S,P}$ is of the same order than the $\Lambda_{\gamma\gamma}$ one and for visible ALP decays in two photons.

On the other hand, the bound in \eq{limitLambdaS} gives a stronger constraint on the effective scale $\Lambda_S$, which now reads $\Lambda_S > 4.7\times 10\, (10^3)$TeV for $m_S\simeq 0.1 (1) {\rm GeV}$, corresponding to a BR of order  ${\rm BR}(Z\to \varphi \gamma)\simeq 1.8\times 10^{-6} (10^{-10})$ respectively.
This bound is consistent with the  upper limits of $10^{-6}$ on the BR from LEP negative searches of this signature \cite{Acciarri:1997im}.
As we can see, these BRs are even larger than the expected ones in $Z\to \gamma \bar{\gamma}$, and thus potentially candidates to the signature in \eq{Zdecay}. Analogous conclusions, with BRs of same order, can be obtained for the pseudoscalar case.

\begin{figure}
\begin{center}
\hspace{0.cm}
\includegraphics[width=0.4\textwidth]{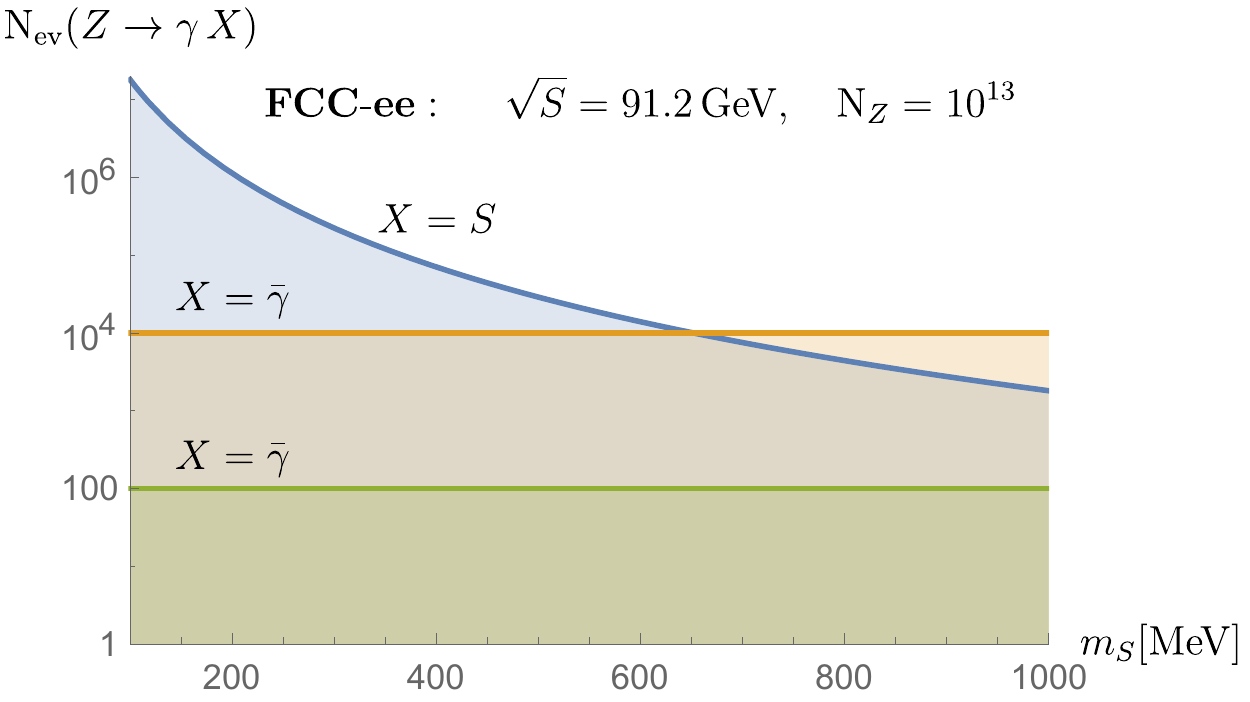}~~~~~~~
\includegraphics[width=0.4\textwidth]{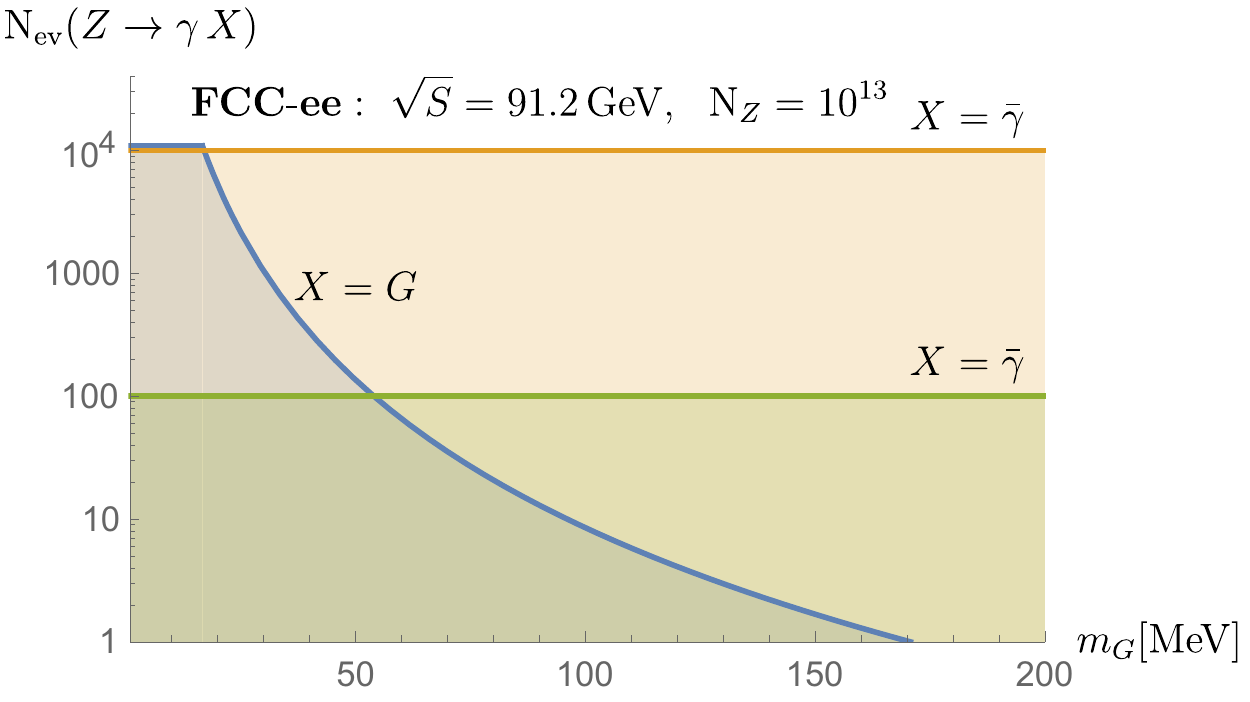}
\caption{{\small Allowed regions (in color) of the number of expected events for the $Z\to \gamma X$ signal,  as a function of the $m_X$ mass in ${\rm MeV}$,  for $X=S$, spin-0 scalar (left-plot) and $X=G$, spin-2 (right-plot). The two horizontal bands correspond to two representative upper bounds of the  ${\rm BR}(Z\to \gamma \bar{\gamma})$ for the ($X=\bar{\gamma}$) massless dark-photon, depending on the choice of free parameters in the dark sector (see section \ref{sec:DP} for more details). These results are based on the assumption of $N_Z=10^{13}$ number of $Z$ bosons events collected at the FCC-ee in the center of mass energy $\sqrt{S}=91.2{\rm GeV}$ at the $Z$ peak.}} 
\label{fig1}
\end{center}
\end{figure}

In the left plot of Fig.\ref{fig1} we summarize the results for the allowed regions (in color) of the number of expected events at $e^+e^-$ colliders, based on the constraints in Eq.(\ref{limitLambdaS}), as a function of the scalar mass $m_S$ in MeV. For comparison the upper bounds on the expected number of events for the massless dark-photon, given by the two horizontal lines, are provided, that correspond to two representative choices for the relevant free parameters in the dark-sector (see section \ref{sec:DP} for more details).
For the results in Fig.\ref{fig1}, we have assumed the largest statistic of $N_Z=10^{13}$ number of $Z$ boson expected to be collected at the FCC-ee collider in the center of mass energy $\sqrt{S}=91.2{\rm GeV}$ at the $Z$ peak.
As we can see, a large number of expected events are possible for a ALP particle, that would allow to study with sufficient precision also angular distributions of the corresponding rates.
 
\subsection{Massive spin-2 particle}
\label{sec:G}
As last example, we  consider the case of a massive spin-2 particle  $X=G$, which is universally coupled to the total energy-momentum tensor $T_{\mu\nu}$ of SM fields. As in the case of a massive graviton, this coupling reads
\bea
L_{\G}=-\frac{1}{\Lambda_{\G}} T^{\mu\nu}G_{\mu\nu}\, ,
\label{Lspin2}
\eea
where $G_{\mu\nu}$ is the field associated to the spin-2 particle $G$.
Since we assume $G_{\mu\nu}$ not to be related to gravitational interactions, we take  the effective scale $\Lambda_{\G}$ as a free parameter, uncorrelated from the Planck mass, and of the order of the ${\rm TeV}$ scale. This scale is reduced to the usual $\Lambda_G^{-1}=\sqrt{8\pi G_N}$ relation in the ordinary case of massless graviton of General Relativity, with $G_N$ the Newton constant.
Since we do not make any hypothesis on the origin of the spin-2 field, we limit ourselves to the linear theory in flat space, avoiding to enter into the issue of a consistent theory of massive spin-2 fields related to the non-linear massive graviton interactions. For the purposes of the present paper the coupling in Eq.(\ref{Lspin2}) is sufficient to generate a finite (thus predictive) contribution at 1-loop for the effective $ZG\gamma$ coupling. Indeed, due to the fact that  $G_{\mu\nu}$ is coupled to the conserved  energy-momentum tensor $T^{\mu\nu}$ of matter fields, the theory is renormalizable against radiative corrections of SM matter fields only, provided the  $G_{\mu\nu}$ is taken as an external field.

The free Lagrangian for the massive spin-2 is given by the usual term of the Fierz-Pauli Lagrangian \cite{Fierz:1939ix} and we do not report its expression here. The corresponding Feynman rules for the $G$ interaction in \eq{Lspin2} can be derived from previous works on massive KK graviton productions in ADD scenarios \cite{Han:1998sg},\cite{Giudice:1998ck}.

Now, we require that the mass $m_{\G}$ of the spin-2 particle is much smaller than the $Z$ mass, but larger than the ${\rm eV}$ scale, in order to avoid the strong constraints from negative searches on the Newton law deviations at short distances \cite{Murata:2014nra}.

The effective $ZG\gamma$ coupling at low energy, is generated  at 1-loop starting from the couplings in \eq{Lspin2}, with $Z,G,\gamma$ external on-shell fields, in which only virtual SM fields run inside. As mentioned above, this contribution is finite due to the conservation of $T_{\mu\nu}$ (at the zero order in $1/\Lambda_G$). This vertex and the corresponding $Z\to \gamma G$ decay has been computed in the context of quantum gravity in large extra dimension scenarios \cite{Allanach:2007ea}, with $G$ the field of a generic massive spin-2 KK excitation of the standard graviton, and for the (massless) graviton in the Einstein theory \cite{Nieves:2005ti}.

Before entering in the discussion of the $Z\to \gamma G$ decay, we analyze the bounds on $m_G$ against the scale $\Lambda_G$, obtained by requiring that $G$ does not decay inside the detector, assumed as in section \ref{sec:ALP} of length $L=10{\rm m}$. Since we are going to discuss a light $G$ which decays into SM particles, as in the ALP case, we restrict the analysis to the range of masses
\bea
    {\rm eV} \lsim m_G \lsim 1\,{\rm GeV}\,.
    \label{Gmass}
\eea
The tree-level total width of a spin-2 particle at rest, decaying into (massless) SM fermion pair $f\bar{f}$, for the Lagrangian interaction in Eq.(\ref{Lspin2}), is given by\cite{Han:1998sg}
\bea
\hat{\Gamma}(G\to \bar{f} f)&=&\frac{m_G^3N_c}{80\pi\Lambda_{\G}^2}
\eea
where $N_c=1$ and $N_c=3$ for leptons and quarks respectively, while the corresponding one for the decay into two massless gauge bosons $V$ is \cite{Han:1998sg} 
\bea
\hat{\Gamma}(G\to VV)=\frac{N_g m_G^3}{40\pi\Lambda_{\G}^2}\,
\eea
where $N_V=1$ and $N_V=8$ for $V=\gamma$ (photons) and $V=g$ (gluons) respectively.

Then, the total width of $G$ in visible sector, corresponding to $m_G=1{\rm GeV}$ can be approximated to
\bea
\Gamma(G\to {\rm visible})\sim 15 \hat{\Gamma}(G\to \gamma\gamma)\, ,
\label{Gwidth1GeV}
\eea
where we neglected all fermion masses, and included channels in two photons, two gluons (assumed here to hadronize in two jets of light mesons), $e^+e^-$,$\mu^+\mu^-$, quark pairs $q\bar{q}$ for $q=u,d,s$.

In order to simplify the analysis, we divide the range of $m_G$ in two regions, below and above the di-muon mass threshold $2m_{\mu}$. In the first region, only the two photon and electron pair channel contribute to the total width. For the second region, we assume the largest value for the total width $\Gamma(G\to {\rm visible})$ corresponding to $m_G=1{\rm GeV}$, where all channels mentioned above contribute, that is a quite good approximation for our estimate.
Then, by requiring that the spin-2 particle does not decay into visible states inside the detector -- unlike the decay into neutrino-pairs which is allowed --  we get an upper bound on $m_G$ versus $\Lambda_G$ as in the ALP case,  namely
\bea
\Lambda_G &\gsim& 36 \left(\frac{m_G}{\rm 100 MeV}\right)^2\, {\rm TeV}\,
\, , ~~~~~~~1{\rm eV} \lsim m_G \lsim  2m_{\mu}
\nonumber\\
\Lambda_G &\gsim& 113 \left(\frac{m_G}{\rm 100 MeV}\right)^2\, {\rm TeV}\,
\, , ~~~~~~2m_{\mu}\lsim m_G \lsim 1{\rm GeV}\, . 
\label{limitMG}
\eea

Further theoretical constraints on this scenario should be imposed on the scale $\Lambda$ that can replace the bounds in \eq{limitMG} for masses below 10 MeV with stronger constraints. In particular, in order to suppress potential large contributions from \textit{Bremsstrahlung} of $G$ in high energy experiments, that would break perturbative unitarity at the TeV energy colliders, we require that $\Lambda_G > {\cal O}(1{\rm TeV})$ for all masses below $10{\rm MeV}$.
Finally, from these results we can see that for a mass range 
${\rm eV} < m_G\sim 10 {\rm MeV}$ we have $\Lambda_G\gsim  1 {\rm TeV}$,
while for $m_G\sim 50(100){\rm MeV}$ we get $\Lambda_G\gsim 28 (113) {\rm TeV}$.

Now, we compute the ${\rm BR}(Z\to \gamma G)$ as a function of the $\Lambda_G$ scale. The corresponding amplitude $M_{\G}$ for the process
\bea
Z(p)&\to& \gamma(k)~ G(q)
\eea
is induced at 1-loop and it is given by \cite{Allanach:2007ea}
\bea
M_{\G}&=& F_{\G}
\varepsilon_Z^{\mu}(p)\varepsilon^{\lambda\rho \star}_{\G}(q)\varepsilon^{\nu\star}(k) V^{\G}_{\mu\lambda\rho\nu}(k,q)
\eea
where  $\varepsilon^{\lambda\rho }_{\G}(q)$ is the polarization tensor of the massive spin-2 field. The $F_{\G}$ is a form factor which is the result of a 1-loop computation. It depends only by SM parameters. Its expression can be found in \cite{Allanach:2007ea} and \cite{Nieves:2005ti} for massive and massless $G$ respectively (with notation $F_h$). The effective vertex $V^{\G}_{\mu\lambda\rho\nu}(p,q)$ is  \cite{Allanach:2007ea}
\bea
V^{\G}_{\mu\lambda\rho\nu}(k,q)&=&
  \left(k_{\lambda} q_{\nu}-(k\cdot q) \eta_{\nu\lambda}\right)
  \left(k_{\rho} q_{\mu}-(k\cdot q) \eta_{\mu\rho}\right)\,+\,\{\lambda\leftrightarrow \rho\}\, .
  \eea
The form factor $F_{\G}$ is \cite{Nieves:2005ti,Allanach:2007ea}
  \bea
  F_{\G}&\simeq  &0.41\frac{\alpha}{\Lambda_G M_Z^2\pi}\, .
  \eea   
After computing the square of the amplitude and summed over all polarizations, mediating by the initial ones, the unpolarized total width in the $Z$ rest frame is
  \bea
  \hat{\Gamma}_{\!\G}=\frac{M_Z^7}{576\pi}\left(7+3r_{\G}\right)\left(1-r_{\G}\right)^5 |F_{\G}|^2\, ,
  \label{GammaZG}
  \eea
where $r_G=m_G^2/M_Z^2$, which, in the small $m_{\G}$ limit, reduces to\footnote{
 Notice that the massless limit of the width in \eq{GammaZG0} differs from the corresponding one for pure massless graviton \cite{Nieves:2005ti}, by a overall factor 7/6, which is due to the sum over polarizations of massive graviton with respect to the massless one. This is due to the known van Dam--Veltman discontinuity in the $m_G\to 0$ limit \cite{vanDam:1970vg}.}
\bea
\hat{\Gamma}_{\!\G}=\frac{7 M_Z^7}{576\pi}|F_{\G}|^2 +{\cal O}(r_G)\, .
\label{GammaZG0}
  \eea
  The result in \eq{GammaZG} is in agreement with the corresponding one in \cite{Allanach:2007ea}.
 Numerically this gives
  \bea
  \hat{\Gamma}_{\!\G}\simeq 2.7 \times 10^{-9} \left(\frac{1 {\rm TeV}}{\Lambda_{\G}}\right)^2 {\rm GeV}\, ,
  \label{GammaZGnum}
  \eea 
  corresponding to a branching ratio
  \bea
      {\rm BR}(Z\to \gamma G) = 1.1 \times 10^{-9} \left(\frac{1 {\rm TeV}}{\Lambda_{\G}}\right)^2\, .
      \label{BRZG}
  \eea
  Finally, by using the results in Eqs.(\ref{limitMG}),(\ref{BRZG}), we find that a viable BR for the signal in \eq{Zdecay} mediated by a long-lived spin-2 particle $G$ in the range  $10^{-12}\lsim {\rm BR}(Z\to \gamma G) \lsim 10^{-9}$ is possible, for a mass range between ${\rm 1eV} < m_G < 50 {\rm MeV}$. For spin-2 masses above 50 MeV scale, the requirement of missing energy signature which is set in the upper bounds in \eq{limitMG}, would exclude the BR above the
  $10^{-12}$ limit.

In the right-plot of Fig.\ref{fig1}, the allowed regions (in color) for the number of expected events at $e^+e^-$ colliders are shown, as a function of the scalar mass $m_G$ in MeV. These bounds are mainly based on the constraints in
Eqs.(\ref{limitMG}) and are based on the number $N_Z=10^{13}$ of $Z$ boson collected at the FCC-ee.
For comparison, the two horizontal lines corresponding to the expected events in the massless dark-photon scenario, for two representative values of dark sector couplings.
The flat dependence of the upper bounds of number events for the spin-2 case, corresponds to the mass-independent lower bound on the corresponding effective scale for $\Lambda_G\gsim 1{\rm TeV}$ coming from negative searches of light spin-2 production at LHC, as explained above, which overseeds the lower bounds in Eqs.(\ref{limitMG}) for spin-2 masses $m_G\lsim 30{\rm MeV}$.
\section{Polarized processes}
\label{sec:pol}
Here we analyze the angular distributions for the decays $Z\to X\gamma$, summed over all polarizations of final states, at fixed polarizations of the $Z$ boson , for the three $X$ scenarios discussed above. The reason to focus on the polarized processes is because the $Z$ boson (on-shell) is always produced polarized at colliders, due to its couplings to SM fermions. We will show in more details this feature in the following, for the particular case of a $Z$ boson production in a resonant $s$-channel at $e^+e^-$ colliders.

In order analyze the polarized $Z$ decays, we need to identify a special direction against which to consider its projections. In this respect, we choose a frame in which the $Z$ is boosted, and identify this direction with the one parallel to the $Z$ 3-momentum $\vec{p}_Z$, that we choose along the $z$-axis, in particular
\bea
p_Z&=&E_Z(1,0,0,\beta)\, .
\label{Zdir}
\eea
where $\beta=\sqrt{1-\frac{M_Z^2}{E_Z^2}}$ is the $Z$ velocity.
In this frame the differential $Z$ decay width $d \Gamma$ reads
\bea
d \Gamma &=& \frac{|M|^2 M_Z^2}{32 \pi E_Z^3(1-\beta z)^2}d z
\eea
where $|M|^2$ is the corresponding (Lorentz invariant) square modulus of the amplitude, $z\equiv \cos{\theta_{\gamma}}$ with $\theta_{\gamma}$ the angle between the $Z$ and the photon 3-momenta. The distributions for the various spin cases $S_X=1,0,2$ in this frame are discussed below.

\vspace{0.5cm}
{\bf Massless dark photon --}
We consider first the case of a $X$ to be a massless dark photon.
We anticipate here that the angular distributions of the photon for the polarized Z decay induced by magnetic and electric dipole moments interactions are the same.

We define the longitudinal $(L)$ and transverse $(T)$ $Z$ polarizations  with respect to the $Z$ momentum in \eq{Zdir} respectively, corresponding to the eigenstates of spin projection along the $z$ axis $J_z=\pm 1$ and  $J_z=0$ respectively. Then, the final result for these distributions  is
\bea
\frac{1}{\hat{\Gamma}}\frac{d \Gamma^{(T)}}{dz}&=&
\frac{3}{4} \left(\frac{M_Z}{E_Z}\right)^5
\frac{1-z^2}{\left(1-\beta z\right)^4}\, ,
\label{dDPT}
\\
\frac{1}{\hat{\Gamma}}\frac{d \Gamma^{(L)}}{d z}&=&
\frac{3}{2} \left(\frac{M_Z}{E_Z}\right)^3
  \frac{\left(\beta-z\right)^2}{\left(1-\beta z\right)^4}
\label{dDPL}
  \eea
  where
  $\hat{\Gamma}$ is the total width  in the $Z$ rest frame given in
  \eq{GammaZDP}.
  
In Eq.(\ref{dDPT}), the distribution $d \Gamma^{(T)}$ for the transverse polarizations  $J_z=\pm 1$ includes the average factor (1/2) over initial polarizations. The angular distributions corresponding to the two transverse polarization states $J_z=\pm 1$ are identical.
As a quick check, we can see that the angular distribution in $Z$ rest frame ($\beta=0$) for the unpolarized process, given by
\bea
\frac{d \Gamma}{d z}|_{\beta=0}&=&\left(\frac{2}{3} \frac{d \Gamma^{(T)}}{d z} +\frac{1}{3} \frac{d \Gamma^{(L)}}{dz }\right)_{\beta=0}=\frac{\hat{\Gamma}}{2}\, ,
\label{Zdistr1}
  \eea
is isotropic, in agreement with known theoretical expectations. Also, by integrating \eq{Zdistr1} at $\beta\neq 0$, the value of the total width in the moving frame $\int_{-1}^1d z \frac{d \Gamma}{d z}=\frac{M_Z}{E_Z}\hat{\Gamma}$ is recovered. 

In the $Z$ rest frame ($\beta\to 0$, $E_Z\to M_Z$), where any direction is equivalent, the angle $\theta_{\gamma}$ is identified here with the angle formed between the directions of photon-momentum and the $z$-axis, the latter being the axis where the $Z$ spin projections have determined values. Then, the corresponding distributions of \eq{dDPT} and \eq{dDPL} in the $Z$ rest frame for the massless dark-photon are
\bea
\frac{1}{\hat{\Gamma}}\frac{d \Gamma^{(T)}}{dz}&=&
\frac{3}{4}\left(1-z^2\right)\, ,
\label{dDPTcm}
\\
\frac{1}{\hat{\Gamma}}\frac{d \Gamma^{(L)}}{d z}&=&
\frac{3}{2}z^2
\label{dDPLcm}
  \eea
We will see in the next section that, due to the $Z$ couplings to electrons, in the resonant production at $e^+e^-$ the $Z$ is mainly produced polarized at rest with transverse polarizations with respect to the beam axis.

\vskip0.3em 
\textit{{\bf Scalar and pseudoscalar --}}
Now, we repeat the same analysis above,  but in the case of $Z$ decays into photon plus a scalar $S$ or a pseudoscalar $P$, in the massless limit.
Since the polarized angular distributions for the scalar and pseudoscalar cases are the same, we will show only one of them as a representative case. Then, the results for these distributions, normalized to the corresponding total width, are
\bea
\frac{1}{\hat{\Gamma}_I}\frac{d \Gamma^{(T)}_I}{d z}&=&
\frac{3}{8} \left(\frac{M_Z}{E_Z}\right)^3
\frac{(1+z^2)(1+\beta^2)-4\beta z}{\left(1-\beta z\right)^4}\, ,
\label{dST}
\\
\frac{1}{\hat{\Gamma}_I}\frac{d \Gamma^{(L)}_I}{d z}&=&
\frac{3}{4} \left(\frac{M_Z}{E_Z}\right)^5
\frac{1-z^2}{\left(1-\beta z\right)^4}
\label{dSL}
\eea
with $\hat{\Gamma}_I$ the total width for $I=S,P$ given in \eq{GammaZS}. The distributions for the two transverse polarizations are the same.
As for the spin-1 case, one can check that in the unpolarized case, the $Z$ the distribution in the $Z$ rest frame is independent by the angle $\theta_{\gamma}$, and that by integrating in $\theta_{\gamma}$ the total width for the unpolarized distribution in \eq{GammaZS} is recovered.

In the $Z$ rest frame and including also the exact $X$ mass effects, we get
\bea
\frac{1}{\hat{\Gamma}_I}\frac{d \Gamma^{(T)}_I}{d z}&=&
\frac{3}{8}\left(1+z^2\right)
\label{dSTcm}
\\
\frac{1}{\hat{\Gamma}_I}\frac{d \Gamma^{(L)}_I}{d z}&=&
\frac{3}{4}\left(1-z^2\right)\, .
\label{dSLcm}
\eea
Notice that these are the same distributions of the $X=S/P$ massless limit.

\begin{figure}
\begin{center}
\hspace{0.cm}
\includegraphics[width=0.4\textwidth]{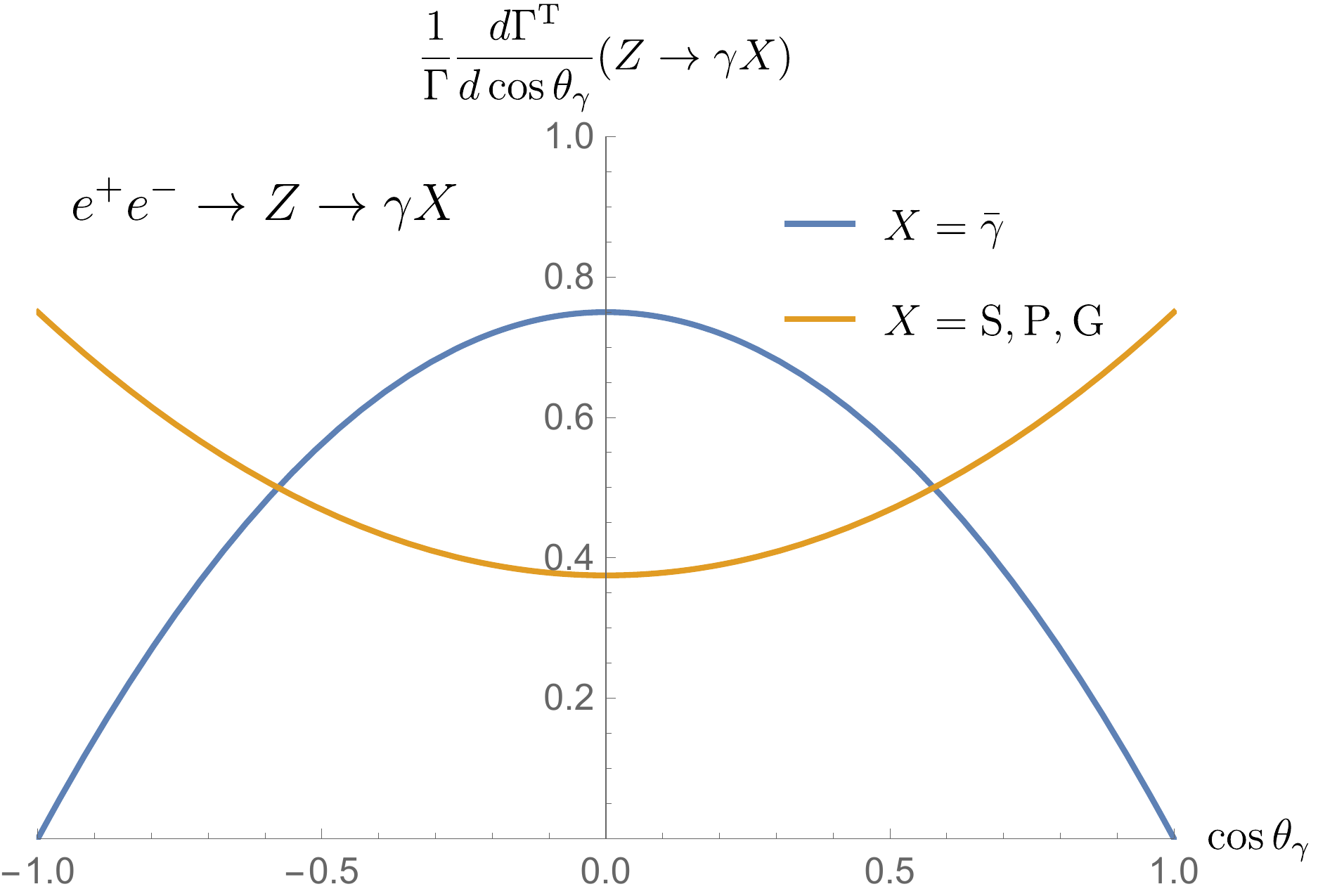}~~~~~~~
\includegraphics[width=0.4\textwidth]{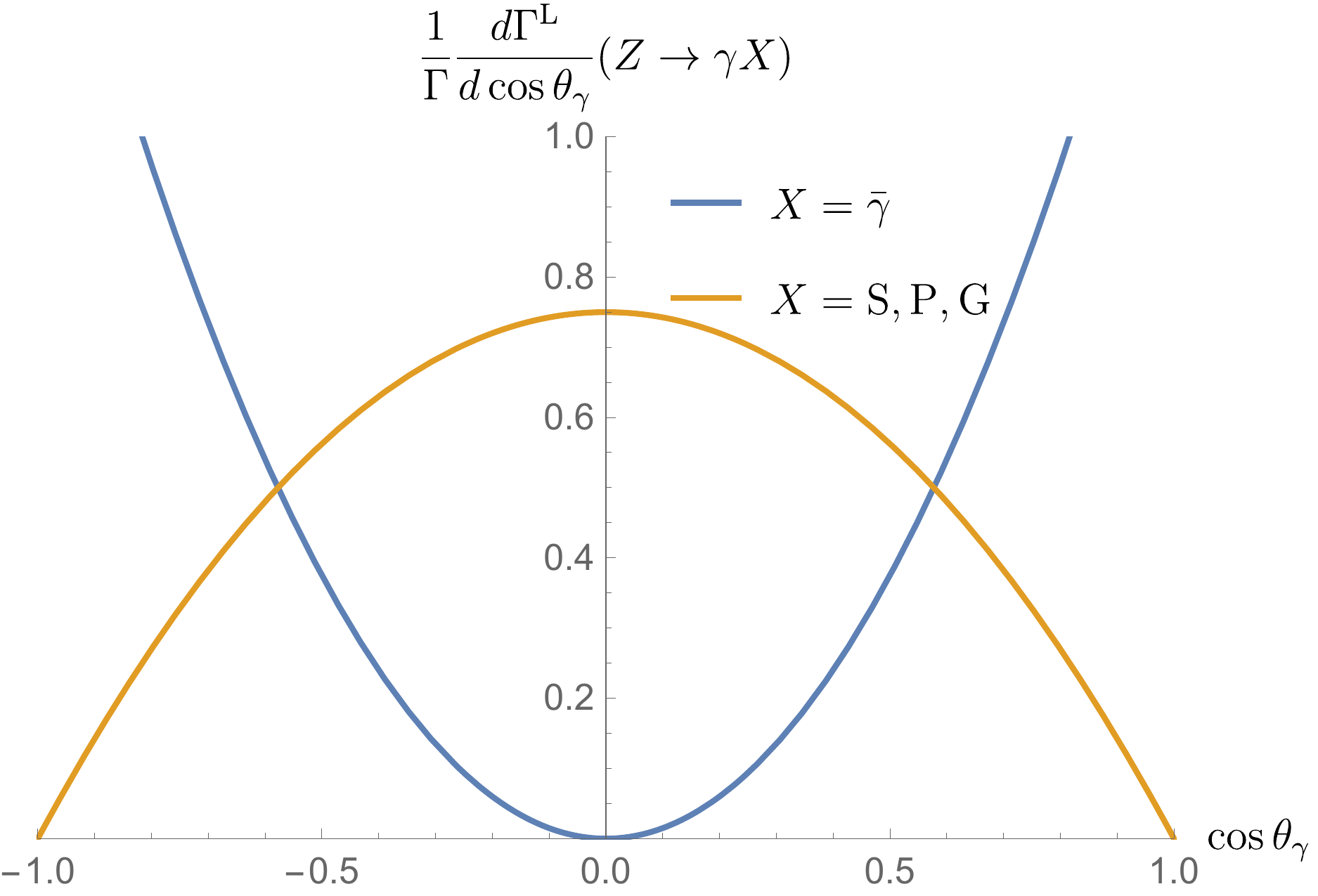}
\caption{{\small Normalized distributions in $\cos{\theta_{\gamma}}$ for the polarized $Z\to \gamma X$ decay in the $Z$ rest frame, with $\Gamma$ the corresponding unpolarized total width, for the scenarios of $X$ as massless dark-photon ($\bar{\gamma}$), scalar/pseudoscalar ($S/P$) and spin-2 ($G$) particles. Here $\theta_{\gamma}$ is the angle between the directions of photon momentum and the $J_z$ spin axis of the $Z$ (see text for details), The distributions of transverse $T$ and longitudinal $L$ polarizations of the $Z$, corresponding to $J_z=\pm 1$ and $J_z=0$ respectively, are shown in the left and right plots respectively. Normalized  angular distributions for a $Z$ produced at rest via $e^+e^-\to Z\to \gamma X$ are shown in the left plot, for the same $X$ scenarios, where there $\theta_{\gamma}$ is the angle between the photon momentum and the beam axis in the center of mass frame. }} 
\label{fig2}
\end{center}
\end{figure}

Remarkably, for the longitudinal and transverse polarizations, the corresponding distributions of the massless spin-1 and spin-0 case are different. These distributions are shown in Fig.\ref{fig2}, including the spin-2 cases $X=G$.

\vskip0.3em 
\textit{{\bf Massive spin-2 --}} Following the same analysis above, we provide below the polarized angular distributions for the spin-2 case in the $Z\to \gamma G$ decay, in the $m_G$ massless limit, normalized to the corresponding total width in the $Z$ rest frame, in particular
\bea
\frac{1}{\hat{\Gamma}_{\!\G}}\frac{d \Gamma^{(T)}_{\!\G}}{d z}&=&
\frac{3}{8} \left(\frac{M_Z}{E_Z}\right)^2
\frac{(1+z^2)(1+\beta^2)-4\beta z}{\left(1-\beta z\right)^4}\, ,
\label{dGT}
\\
\frac{1}{\hat{\Gamma}_{\!\G}}\frac{d \Gamma^{(L)}_{\!\G}}{d z}&=&
\frac{3}{4} \left(\frac{M_Z}{E_Z}\right)^4
\frac{1-z^2}{\left(1-\beta z\right)^4}\, ,
\label{dGL}
\eea
where the total width $\hat{\Gamma}_{\!\G}$ is given in \eq{GammaZG0}. The angular distributions for the two transverse polarizations are identical.
As we can see from these results, the angular distributions of $Z$ for the spin-2 case have the same functional dependence by $\theta_{\gamma}$ of the corresponding scalar/pseudoscalar ones at fixed polarizations, see Eqs.(\ref{dST}),(\ref{dSL}). They only differ in the boosted frame by different powers of $M_Z/E_Z$ in the overall coefficients. This equivalence holds only in the massless limit.

Below we provide the expressions for the angular distributions in the $Z$ rest frame and by retaining the exact mass dependence in $m_{\G}$, in particular
\bea
\frac{1}{\hat{\Gamma}_{\!\G}}\frac{d \Gamma^{(T)}_{\!\G}}{d z}&=&
\frac{3}{8} \frac{\left(1+z^2\left(1-2\delta_{\G}\right)+2\delta_{\G}\right)}{1+\delta_{\G}}
\label{dGTcm}
\\
\frac{1}{\hat{\Gamma}_{\!\G}}\frac{d \Gamma^{(L)}_{\!\G}}{d z}&=&
\frac{3}{4} \frac{\left(1-z^2\left(1-2\delta_{\G}\right)\right)}{1+\delta_{\G}}\, ,
\label{dGLcm}
\eea
where $\delta_{\G}=\frac{3}{7}r_{\G}$.
As we can see, spin-2 $X$ mass corrections are expected to break the equivalence between the spin-0 and spin-2 angular distribution (valid only in the $X$ massless limit) by terms of order $r_{\G}$, which in our case are smaller than $10^{-4}$. The corresponding angular distributions for the spin-2 in the $Z$ rest frame are plotted in Fig.\ref{fig2} (in the massless case).


\section{Z decays at $e^+e^-$ colliders}
\label{sec:ee}

In this section we will analyze the photon angular distributions coming from the $Z$-resonant process $e^+e^-\to Z\to \gamma X$ at the $Z$ peak.
We will show that these distributions can be easily obtained from a particular linear combination of polarized $Z$ distributions analyzed above. This approach has the advantage to avoid the computation of the scattering cross section $e^+e^-\to Z\to \gamma X$. These results can also be applied to any final state.

In the center of mass frame of $e^+e^-$, the beam axis identifies a special direction, that we choose to be our third- or $z$-axis.  In this frame, we choose the initial momenta along the beam direction, namely $p_{e^-}=(E,0,0,E)$ and $p_{e^+}=(E,0,0,-E)$, where $E=\sqrt{S}/2$ is the center of mass energy (we neglect the electron mass). The transverse and longitudinal $Z$ polarizations for a $Z$ at rest can now be identified with respect to the beam axis.

In this frame,  we define the two transverse $Z$ polarizations vectors, for a $Z$ at rest, as 
\bea
\varepsilon_Z^{\mu\, (\pm)}&=&\frac{1}{\sqrt{2}}\left(0,1,\pm i,0\right)\, ,
\eea
while for the longitudinal one, for a $Z$ at rest, we have 
\bea
\varepsilon_Z^{\mu\, (L)}&=&\left(0,0,0,1\right)\, .
\eea
In the case of a frame with boosted $Z$ along the beam direction, in which the $Z$ comes out with a velocity $\beta =k_Z/E_Z$, with $k_Z$ and $E_Z$ its momentum and energy respectively, the corresponding results for the longitudinal polarization generalize to 
\bea 
\varepsilon_Z^{\mu\, (L)}&=&\frac{1}{M_Z}\left(k_Z,0,0,E_Z\right)\, .
\eea

Then, concerning our final state, we identify the angle $\theta_{\gamma}$ as the angle formed between the direction of the outgoing photon momentum and the initial electron momentum $\vec{p}_{e^-}$, in particular for the photon 4-momentum we have
\bea
k_{\gamma}=\frac{E}{2}\left(1,\sin{\theta_{\gamma}}\cos{\phi_{\gamma}},\sin{\theta_{\gamma}}\sin{\phi_{\gamma}},\cos{\theta_{\gamma}}\right)\, ,
\label{kphoton}
\eea
with $\phi_{\gamma}$ the corresponding photon azimuthal angle.

Now, we can extract the $\cos{\theta_{\gamma}}$ distributions of the final photon in $e^+e^-\to Z\to \gamma X$, by using a linear combination of the same polarized $Z$ angular distributions discussed in previous section, provided the $\theta$ angle appearing in the $z=\cos{\theta}$ distributions in \eqs{dDPT}{dGL} is identified with the $\theta$ angle defined in \eq{kphoton}. In this linear combination, each contribution of the $Z$ polarization $\varepsilon_Z^{(\lambda)}$ to the width should be multiplied by a polarization-weight coefficient $0\le C_Z^{(\lambda)}\le 1$ (where $\sum_{\lambda=\pm,L}C_Z^{(\lambda)}=1$), corresponding to the $Z$ production in resonance $e^+e^-$ collision.

We find these coefficients  $C_Z^{(\lambda)}$ by performing the matching between the resonant $e^+e^-\to Z\to X_f$ cross section (with $X_f$ a generic final state) in the Breit-Wigner approximation, against the decay width of a polarized on-shell $Z$ boson. These coefficients are universal, since they depend only by the initial states, which in this case are the $e^+e^-$ from which the $Z$ has been created. Therefore, these results could be applied to any final state.

In general, for $e^+e^-$ collisions, a generic distribution of final states $d\Gamma_f$ reads 
\bea
d \Gamma_f(e^+e^-\to Z\to X_f)&=& C_Z^{+} d \Gamma^{+}_f +C_Z^{-}d \Gamma^{-}_f\, + C_Z^{L} d \Gamma^{L}_f\, ,
\label{dgammapol}
\eea
where $d \Gamma^{\pm}_f$ ($d \Gamma^{L}_f$) stand for the corresponding transverse (longitudinal) polarized distributions of the $Z\to X_f$ decay and $C_Z^{\pm,L}$ the corresponding polarization weights. For a $Z$ boson at rest, we have
\bea
C_Z^{\pm}&=& \frac{1}{2}\left(1\mp \frac{2g_V^eg_A^e}{(g_V^e)^2+(g_A^e)^2}\right),~~~C_Z^{L}={\cal O}(m_e/M_Z)\, ,
\label{Cpm}
\eea
with $m_e$ the electron mass.
As we can see from the above results \eq{dgammapol}, the contribution of the longitudinal polarization $\varepsilon_Z^{L}$ is strongly suppressed and vanishing in the limit of vanishing electron mass. This means that the $Z$ boson produced in resonance at $e^+e^-$ comes out mainly transverse polarized with respect to the beam direction. This is a well known result that can be easily understood in terms of chirality arguments and angular momentum conservation.

The relation in \eq{dgammapol} can be applied to all kind of distribution of final states. In particular, it reproduces the well known result of angular distributions of fermion pair production $e^+e^-\to Z\to f \bar{f}$ in the $Z$ resonant region at the peak, including the contribution to the forward-backward (FB) asymmetry.

In general, for a boosted frame in which the resonant $Z$ is produced with speed $\beta$ along the beam direction, the polarization coefficients $C^{\pm}$ read
\bea
C^{\pm}&=& \frac{1}{2}\left(1\mp \frac{2g_V^eg_A^e}{(g_V^e)^2+(g_A^e)^2}
\frac{(1-\beta)^2}{1-\beta^2}\right)\, .
\label{Cpmboosted}
\eea
These results could be also generalized to a resonant $Z$ production at hadron colliders via quark-antiquark annihilation, provided in \eq{Cpmboosted} $g_{V,A}^e$ are replaced with the corresponding $g_{V,A}^u$ and $g_{V,A}^d$ couplings to up and down quarks respectively.

The term proportional to $(\mp)$ coefficient in \eqs{Cpm}{Cpmboosted}, is responsible of parity violating contributions. We find that, in all spin cases analyzed here for the  $Z\to \gamma X$ process, the two angular distributions $\frac{d\Gamma^{+}}{dz}=\frac{d\Gamma^{-}}{dz}$ are the same for all processes. This means that the $C^{\pm}$ polarization coefficients enter in the combination of $C^++C^-=1$ for  a $Z\to \gamma X$ decay produced in resonance at $e^+e^-$ colliders. This is due to the fact that, the $Z$ bosonic effective vertices discussed above do not introduce any parity violating contributions when the $Z$ is produced from an unpolarized $e^+e^-$ collider.

In conclusion, the photon angular ($\theta_{\gamma}$) distributions, coming from the resonant $Z$ boson produced in $e^+e^-$, are just given by the $d \Gamma^{T}_f/d z$ expressions reported in \eqs{dDPT}{dGL}, and are shown in the left plot of Fig.\ref{fig2} for the various $X$ scenarios.

From these results we could see that a  massless dark-photon signature is indeed characterized by a central photon, produced at large angles $\theta$ with respect to the beam, while it is vanishing in the FB directions
$(\theta=0,\pi)$. On the other hand, for the spin-0 and spin-2 cases the photon will be mainly emitted in the FB directions. This is also in agreement with results on photon angular distributions in the KK gravitons emission in the massless limit \cite{Allanach:2007ea}. This behaviour can be easily understood by angular momentum conservation. Due to the conservation of chirality in the $Z$ couplings to initial $e^+e^-$ states, the total angular momentum $J_Z$ along the beam axis could be $J_Z=\pm1$. On the other hand, at $\theta=0,\pi$ where orbital angular momentum vanishes, the two final photon states can have either $J_Z=2,0$, but not $J_Z=0$. This forces the angular distribution rate to vanish at $\theta=0,\pi$ as shown in the left plot of Fig.\ref{fig2}. This conclusion does not hold for the $Z$ decay into a spin-0 or spin-2 particles accompanied by a photon, for which the total $J_Z=1$ is possible at $\theta=0$, leaving to a non-vanishing distribution rate in the FB directions.

These results suggest that from the study of the photon-angular distributions of the $Z\to \gamma X$ decay at an unpolarized $e^+e^-$ it would be possible to disentangle the (massless) $J^P=1^{-}$ nature of the $X$ particle from the other $J^P=0^{-},2^{-}$ hypothesis.
In particular, following the results of \cite{Cobal:2020hmk}, based on the search for dark-photon signal $Z\to \gamma \bar{\gamma}$ at hadronic and $e^+e^-$ future colliders, the lower bound $N$ for the expected and observed number of signal events needed to exclude the $J^P=0^{-}$ test hypothesis under the $J^P=1^{-}$ assumption, is respectively $N=6$ and $N=17$ at the 95\% C.L.
Combining these results with the expected number of allowed events in Fig.\ref{fig1} for the massive ALP particles with these characteristics, we can conclude that there should be sufficient number of viable events to disentangle the spin-1 versus spin-0 hypothesis at the future linear FCC-ee collider. 

On the other hand, it would not be possible to distinguish the $J^P=0^{-}$ from the $J^P=2^{-}$ signals, even by using the facility of polarized beams in the linear $e^+e^-$ colliders. The latter can offer in general great advantages in  enhancing the sensitivity to new physics signal events against the SM background, as for instance in the search of scalar lepton partners in SUSY inspired models \cite{Baum:2020gjj}.
In our particular case, in a polarized $e^+e^-$ collider one can in principle select the polarization weights $C^{(\pm)}$ of the transverse polarizations $\varepsilon_Z^{(\pm)}$ of the $Z$ boson in the resonant $Z$ boson production at $e^+e^-$, offering an extra tool in addition to the unpolarized circular $e^+e^-$ colliders where these coefficients are fixed. 

However, in the $Z\to \gamma X$ decay, due to the kind of interactions involved, the angular distributions of the $X$ spin-0 and spin-2 are identical (in the $X$ massless limit) for all the two $Z$ transverse polarizations (with respect to the beam axis) $\varepsilon_Z^{(+)}$ and  $\varepsilon_Z^{(-)}$. 
Then, the only way to disentangle the $X$ spin-0 against the spin-2 is by using the sensitivity in the angular distribution to the $X$ mass $(m_X)$ effects
(cf. Eqs.(\ref{dSTcm}) and (\ref{dGTcm})), but this task can be also achieved by an unpolarized $e^+e^-$ collider.

Concerning the sensitivity of the angular distributions of the spin-0 versus the spin-2 mass, we can see that this is quite hopeless if the mass of an invisible $X$ of spin-0 or spin-2 is restricted to be below the $1{\rm GeV}$ scale. On the other hand, for larger $X$ masses, the requirement to behave as an invisible particle in the detector sets quite strong constraints on the associated scale for masses larger than $1{\rm GeV}$, see BR allowed regions in Fig.2, thus resulting in a too small number of (viable) events needed to analyze the angular distributions.

 \section{Conclusions}
 We analyzed the  decays of the $Z$ boson into $Z\to \gamma X$ with $X$ a long-lived light dark boson, assumed to behave as missing energy in the detector. We discussed three potential scenarios for $X$ based on their spin origin: a massless or ultralight dark photon for the spin-1, an ALP in both scalar and pseudoscalar cases for the spin-0, and a light spin-2 particle. For the spin-0 and spin-2 scenarios, the masses are assumed to be in the range of $[100 {\rm MeV}-1 {\rm GeV}]$ for the ALP, and  $[1 {\rm eV}-1 {\rm GeV}]$ for the spin-2. Moreover, we required that the ALP and spin-2 particles do not decay inside the detector. We show that for these scenarios the largest BRs could be in the observable range of $10^{-12} \lsim {\rm BR}(Z\to \gamma X) \lsim 10^{-6}$, depending on the spin and allowed values of the corresponding effective scales. All these BRs are in the ballpark of sensitivity range of future $Z$ factories at $e^+e^-$ colliders, like for instance the FCC-ee facility, with its projected production of $10^{13}~Z$ bosons \cite{Abada:2019zxq}.

These scenarios have in common the same signature, characterized by a mono-chromatic photon plus an almost neutrino-like missing energy. In  case this signature should be discovered, a spin test to discriminate about the spin-1 dark photon origin against the spin-0/2 ones is proposed. Due to the fact that the $Z$ boson is always polarized when is resonantly produced  at $e^+e^-$ colliders,  we show that the spin-1 nature of $X$ could be disentangled from the spin-0 and spin-2, by analyzing the angular distribution of the mono-chromatic photon. The massless dark-photon signature is indeed characterized by a photon mainly produced central and at large angles with respect to the $e^+e^-$ beam axis. On the other hand, for the spin-0 and spin-2 cases (that have the same angular distributions) the mono-chromatic photon is mainly expected along the forward/backward directions.

 In conclusion, due to the clean environment of the FCC-ee facility, together with its expectations on the high statistics of the $Z$ bosons collected, the rare
 $Z\to \gamma X$ decay could be a golden place to search for a light $X$ dark boson at the FCC-ee, offering also the possibility of untangling its spin origin.

 \section{Appendix}
 We provide here the expression of the $C_{1-3}$ and $C_E$ coefficients appearing in Eqs.(\ref{LeffMD}),(\ref{LeffED}) for the effective $Z\gamma \bar{\gamma}$ interactions, as a function of the $d_{M,E}^f$ coefficient in \eq{dipole}. By matching the on-shell amplitude for the $Z\to \gamma \bar{\gamma}$ process -- as obtained by using the effective Lagrangian in
 Eqs.(\ref{LeffMD}),(\ref{LeffED}) --- with the corresponding one obtained by the one-loop computation with the insertion of the dipole-operators in \eq{dipole}, we obtain \cite{Fabbrichesi:2017zsc}
\bea
C_1&=&-\sum_f \frac{ d_M^f X_f}{4\pi^2}\Big( 5+2B_f+2C_f\left(m_f^2+M_Z^2\right)\Big)\, ,\nn \\
C_2&=&-3\sum_f \frac{d_M^f X_f}{4\pi^2} \Big(2+B_f\Big)\, , \nn \\
C_3&=&2\sum_f  \frac{d_M^f X_f}{4\pi^2} \Big(4+2B_f+C_fM_Z^2\Big)\, .
\label{CE}
\eea
and
\bea
C_E&=& \sum_f \frac{d^f_E  X_f}{4\pi^2} \Big(3 +B_f+2m^2_fC_f\Big)\, ,
\label{CM}
\eea
where $X_f\equiv \frac{m_f}{M_Z} N_c^f g_A^f  Q_f e_D$, with $m_f$ the mass, $g_A^f$ the $Z$ axial coupling, $Q_f$ the EM charge of SM fermions $f$ in units of $e$, and  $N_c=1 (3)$ for leptons (quarks). The sum over the index $f$ runs over all EM charged SM fermions.
The $B_f$ and $C_f$ terms are defined as
\bea
B_f&\equiv &{\rm Disc}[B_0(M_Z^2, m_f, m_f)], \nn \\
C_f&\equiv & C_0(0, 0, M_Z^2, m_f, m_f, m_f)\, ,
\eea
where $B_0$ and $C_0$ are the scalar two- and three-point Passarino-Veltman functions, respectively (see~\cite{Passarino:1978jh} for their explicit expressions), and ${\rm Disc}[B_0]$ stands for the discontinuity of the $B_0$ function. These terms are both finite functions which can be evaluated numerically, for example, by using the {\tt Package~X}~\cite{Patel:2015tea}. 
\section{Acknowledgments}
We thank M. Fabbrichesi, L. Marzola, B. Mele, and  A. Urbano, for useful discussions. EG is affiliated to the Institute for Fundamental Physics of the Universe (IFPU), Trieste, Italy. EG thanks the Department of Theoretical Physics of CERN  for its kind hospitality during the preparation of this work.

\bibliography{Zdark}    



\end{document}